\documentclass[twocolumn]{aastex62}

\usepackage{times}
\usepackage{graphicx}
\usepackage{epstopdf}
\usepackage{url}
\usepackage{amsmath,amssymb}

\newcommand{\Msun}{{\rm M}_\odot}

\def\ergscm2 {erg\,s$^{-1}$cm$^{-2}$}

\def\cm2 {cm$^{-2}$}
\newcommand{\lsim}{\mathrel{\rlap{\lower 3pt \hbox{$\sim$}} \raise 2.0pt \hbox{$<$}}}
\newcommand{\gsim}{\mathrel{\rlap{\lower 3pt \hbox{$\sim$}} \raise 2.0pt \hbox{$>$}}}

\shorttitle{\textsc{AIC in AGNs}}
\shortauthors{\textsc{ Perna et al.}}

\begin{document}

\title{ACCRETION-INDUCED COLLAPSE OF NEUTRON STARS IN THE DISKS OF ACTIVE GALACTIC NUCLEI}

\author{Rosalba Perna}
\affiliation{Department of Physics and Astronomy, Stony Brook University, Stony Brook, NY 11794-3800, USA}
\affiliation{Center for Computational Astrophysics, Flatiron Institute, New York, NY 10010, USA}

\author{Hiromichi Tagawa}
\affiliation{Astronomical Institute, Graduate School of Science, Tohoku University, Aoba, Sendai 980-8578, Japan}

\author{Zolt\'an Haiman}
\affiliation{Department of Astronomy, Columbia University, 550 W. 120th St., New York, NY, 10027, USA}

\author{Imre Bartos}
\affiliation{Department of Physics, University of Florida, PO Box 118440, Gainesville, FL 32611, USA}

\begin{abstract}
The disks of active galactic nuclei (AGNs) have emerged as a rich environment for the evolution of stars and their compact remnants. The very dense medium favors rapid accretion, while torques and migration traps enhance binary formation and mergers. Both long and short gamma-ray bursts (GRBs) are hence expected.
We show that AGN disks constitute 
an ideal environment 
for another interesting phenomenon: the accretion induced collapse (AIC) of neutron stars (NSs) to black holes (BHs). Rapid accretion in the dense disks can cause NSs to grow to the point of exceeding the maximum mass allowed by their equation of state. 
General relativistic magnetohydrodynamical simulations have shown that electromagnetic signatures are expected if the NS is surrounded by a mini-disk prior to collapse, which then rapidly accretes onto the BH,  and/or if the 
NS is highly magnetized, from reconnection of the magnetosphere during collapse. Here we compute the rates of AICs and their locations within the disks for both isolated NSs, and for (initially stable) NSs formed from NS-NS mergers. We find that the global AIC rates are $\sim 0.07-20$~Gpc$^{-3}$~yr$^{-1}$, and we discuss their observable prospects and signatures as they emerge from the dense disk environments.
\end{abstract}


\section{Introduction}

The accretion disks of Active Galactic Nuclei (AGNs) have emerged as a rich environment for compact objects to interact, grow by accretion of gas, form binaries, and merge. Interest in these environments has been fueled in the last few years 
by some of the recent discoveries by the LIGO and Virgo observatories \citep{2015CQGra..32g4001L,2015CQGra..32b4001A,2018arXiv181112907T,2020arXiv201014527A}.
Some of the compact objects discovered via gravitational waves at the time of their mergers have revealed surprising properties. In particular, the detection of a binary black hole (BH) merger with one of the two components above the traditional range for the pair instability gap \citep{Abbott2020highM}, and another with one of the objects in the lower mass gap ($\sim 2.2-5 M_\odot$, \citealt{Abbott2020lowM}), can both be explained in an AGN environment \citep{Yang2020,Samsing2020,Tagawa2020b,Tagawa2021_Ecc}.
While alternative scenarios remain possible \citep{safarzadeh2020,Renzo2020,Belczynski2020}, AGN disks provide a natural environment for neutron stars (NSs) and BHs to grow and merge multiple times within the AGN lifetime, hence yielding masses that may be larger than predicted in the 'unaided', common formation scenarios of standard stellar evolution (e.g. \citealt{MacKernan2012,Bartos2017,Secunda2019,Yang2019,Yang2020,Tagawa2020}). 

Of particular interest to this work is the growth of NSs in the dense AGN disks by an amount that brings them above the threshold to collapse to a BH. This interesting phenomenon of Accretion Induced Collapse (AIC) of a NS into a BH has been studied theoretically and numerically \citep{Shibata2000,Shibata2003,Baiotti2005,Baiotti2006,Lehner2012,Dionysopolou2013,Palenzuela2013}, and associated with various types of electromagnetic transients (see \S~2 and references therein); however,  there has been no astrophysical context in which such an event has been observed. 
Furthermore, the possible prevalence of this phenomenon, in particular the expected overall event rate, has not previously been quantitatively assessed.

Here we show that AGN disks are ideal environments for creating AICs via multiple channels (\S~2). We thus compute the rates of these events (\S~3), and discuss their observability as transients (\S~4).  We summarize and conclude in \S~5.

\section{Electromagnetic radiation from Collapsing supramassive neutron stars}

The collapse of a NS to a BH in vacuum has been the subject of several numerical simulations \citep{Shibata2000,Shibata2003,Baiotti2005,Baiotti2006,Lehner2012,Dionysopolou2013,Palenzuela2013}.   Some focus has been given to the extraction of the GW signal emitted during the collapse, which has been found to be very weak and only detectable by Advanced LIGO and Virgo for galactic sources. 
Additionally, and of relevance for this work,
NSs collapsing to BHs have been of interest also in light of possible electromagnetic (EM) signatures 
associated with the collapse. In particular, they have been suggested as a possible mechanism for both long \citep{Vietri1999} and short GRBs \citep{MacFadyen2005,Dermer2006}, as well as progenitors of the Fast Radio Bursts (FRBs, \citealt{Falcke2014}) discovered by \citet{Lorimer2007}.

In addition to these ideas which apply to NSs collapsing in vacuum
(and which will be discussed more below), the environments of AGNs, being especially dense, can create additional conditions for the generation of EM counterparts since the NS collapse occurs under the presence of matter.
Full 3D general relativistic simulations of the AIC of a NS to a BH were performed by \citet{Giacomazzo2012} under non-vacuum conditions. In particular, their simulations investigated the effect of a torus of matter surrounding the NS as it collapses to a BH.  In all the models they studied, they found that a fraction  of the mass surrounding the NS prior to collapse remains outside of the BH, and, immediately following the collapse, it rapidly accretes onto the BH.
The accretion is triggered by the spacetime dynamics. The accretion 
rates were found to be on the order of $\sim 10^{-2}M_\odot$~s$^{-1}$, comparable to those inferred for short GRBs. 

These high accretion rates onto the newly formed BH can however be achieved if there is sufficient mass in the close vicinity of the NS at the time of collapse.  Albeit noting the astrophysical possibility of NSs surrounded by large and massive envelopes as a result of special evolutionary pathways \citep{TZ1975}, here we study NS evolution within the specific environment of an AGN disk. To such an effect, we will
need to consider the amount, spatial distribution, and angular momentum of mass in the AGN disk available for accretion by the NS, and the corresponding expected accretion rates.  
This will be quantitatively evaluated in \S~3.4 below. 

In addition to the above mechanism, other ideas proposed for EM counterparts from AICs of NSs to BHs do not necessarily require a reservoir of material outside the collapsing NS. In
the AIC model proposed by \citet{Vietri1998, Vietri1999},  mass transfer onto a NS (either by fallback after a supernova explosion or as a result of accretion from a companion star) can turn it into a supramassive NS (SMNS). SMNSs are rotating NSs with masses larger than the maximum non-rotating value allowed by their equation of state. Such models exist as solutions for all known equations of state at nuclear densities (see e.g. \citealt{Cook1994,Salgado1994}). A key element to the 
\citet{Vietri1998} model is that not all the NS mass will promptly collapse to a BH; the outermost layers, endowed with the largest centrifugal acceleration, remain behind, and hence the end state is  a configuration of a BH surrounded by an equatorial belt in centrifugal equilibrium.  
However, this key assumption has been challenged by  general relativistic (GR) numerical simulations. 
Three-dimensional GR simulations of the collapse of uniformly rotating SMNSs 
by \citet{Shibata2000}  and 
\citet{Baiotti2005}, with polytropic equations of state, have shown that no substantial disk is left outside of the BH event horizon upon collapse. 
Since numerical simulations have only dealt with simplistic equations of state (EoS), the question of the existence of a remnant disk was more generally addressed by \citet{Margalit2015} via an analysis of the  equilibrium hydrostatic configuration of SMNSs prior to the collapse, to assess whether they possess any mass with specific angular momentum exceeding that of the last stable orbit around the nascent BH. By sampling a wide range of NS EoS, they found that disk formation is indeed possible, but not in the parameter space of EoS consistent with NS observations to date. More recently, a similar analysis was conducted by \citet{Camelio2018}. While 
their results are in broad qualitative agreement with those of \citet{Margalit2015}, their simulations found a larger parameter space (for the NS EoS) allowing disk formation, albeit a tenuous one, with mass $10^{-8}- 10^{-7} M_\odot$. 
Based on these analyses, it appears that  significant  energy extraction based on a torus left behind by the collapse is unlikely \footnote{Note that a relatively massive torus is found in numerical simulations (e.g. \citealt{Duez2006,Ciolfi2017}) of the collapse of magnetized, hypermassive NSs (i.e. NSs which are differentially rotating). However, old NSs are expected to have settled in a solid-body rotation.}. 

On the other hand, GR simulations in resistive magnetohydrodynamics with increasing degree of sophistication \citep{Lehner2012, Dionysopolou2013,Palenzuela2013} have shown that the collapse of magnetized NSs is accompanied by a strong electromagnetic emission. This arises due to the fact that, as the NS collapses to a BH, the formation of the event horizon hides most of the matter and radiation, but not the NS magnetosphere\footnote{This is as a result of the no-hair theorem which in this context has however been questioned by some \citep{Lyutikov2011}.}, which thus experiences a violent disruption and reconnection outside the horizon.
This generates strong currents and intense electromagnetic emission. 
\citet{Palenzuela2013} estimated the total radiated energy to be
\begin{equation}
E_{\rm rad}\approx 1.6\times 10^{41} \left(\frac{B_p}{10^{12}~{\rm G}} \right)^2~{\rm erg}\,,
    \label{eq:Erad}
\end{equation}
where $B_p$ is the dipolar field. 
The bulk of this radiation is emitted around the time of the collapse, i.e. within the free-fall time scale 
$\tau_{\rm ff}=0.04(R_{\rm NS}/10{\rm km})^{3/2}(M_{\rm NS}/2.3M_\odot)^{-1/2}$s
\citep{Baiotti2005,Lehner2012},
where $R_{\rm NS}$ indicates the radius of the NS and $M_{\rm NS}$ its mass.
The electromagnetic luminosity is found to appear as a brief pulse of duration $t_{\rm pulse}\approx$ a few tens of  millisecond \citep{Lehner2012,Palenzuela2013}, and its magnitude is hence on the order of
\begin{equation}
L_{\rm EM}\approx 1.6\times 10^{44} \left(\frac{B_p}{10^{12}~{\rm G}} \right)^2\left( \frac{t_{\rm pulse}}{{1~{\rm ms}}}{}\right)^{-1}~{\rm erg}~{\rm s}^{-1}\,.
    \label{eq:Lrad}
\end{equation}
The spectrum of the emitted radiation is notoriously difficult to calculate; some estimates were however made by \citet{Falcke2014} using a basic relativistic radiation curvature model \citep{Gunn1971}. Radiation is emitted at a typical frequency 
\begin{equation}
\nu_{\rm curv} \simeq 7.2~\gamma^3~r_{10}^{-1}~{\rm kHz}\,,    
\end{equation}
where $\gamma$ is the Lorentz factor of the electrons and positrons. For an energy distribution 
${\rm d}N_e(\gamma)/{\rm d}\gamma\propto \gamma^{-p}$ as in shock acceleration, \citet{Falcke2014} estimated the electrons to have a minimum Lorenz factor $\gamma_{\rm min}\gtrsim 175$, and a powerlaw index $p\gtrsim 2.1$, hence making the radiation peak in the GHz frequency range. The characteristics of this emission (both the timescales and the waveband) led \citet{Falcke2014} to suggest that the collapse of a SMNS into a BH could be the progenitor of an FRB. Given the uncertain precise value of the fraction of the radiated luminosity 
which is emitted in the radio, we use as a reference radio luminosity  the value $L_{\rm R}=\eta_{R} L_{\rm EM}$, where $\eta_{\rm R}\approx 0.26$
matches the reference emitted radio power estimated by \citet{Falcke2014}. 

We note that the magnitude of the emission scales with the square of the magnetic field, and hence supramassive magnetars are more likely to be stronger emitters. NSs with very strong fields 
$\gtrsim 10^{14}$~G are born as a fraction of NSs from massive star collapse
\citep{Fryer2004}. 
For Galactic NSs, whose birth rate is dominated by the ones born in supernova explosion, the magnetar fraction is on the order of 10\% \citep{Muno2008}. Massive stars evolving in AGN disks may have a higher fraction of magnetars, since high magnetic fields can be generated via dynamo processes, if the rotation rates are sufficiently high \citep{Raynaud2020}. Recent work \citep{Jermyn2021} has shown that massive stars in AGN disks do evolve rotating very rapidly.
Since the "FRB-like" radiation is emitted around the time of the collapse, while any "GRB-like" radiation (if matter  surrounds the NS at collapse) is emitted later when the accretion-powered outflow becomes optically thin, AICs in AGN disks may appear as FRBs immediately followed ($\sim $~seconds timescale) by a short GRB (see however Sec.~4 for a discussion of the observability of this type of event).

As a separate channel from isolated star evolution, highly magnetized NSs are also expected to be a common outcome of NS mergers (NSM)  which leave behind a stable NS \citep{Giacomazzo2013,Zrake2013,Giacomazzo2015}. The 
spin down time of the post-merger magnetar is relatively quick, $\sim 10^6-10^7$~sec, as a result of losses by magnetic dipole radiation and by gravitational waves, with the relative contribution dependent on the magnetic field strength, initial spin period, and  magnitude of the mass quadrupole moment of the post-merger magnetar \citep{DallOsso2015}. The rapid spin down of the NS ensures that subsequent accretion can proceed effectively, unhindered by propeller effects \citep{Illarionov1975}.

The fraction of NSM which leave behind a stable NS depends on the equation of state of the NS. In the range of EoS studied by 
\citet{Piro2017}, this fraction was found to vary from 0.1\% to 99.6\%. Additional constraints which used also EoS information from GW170817 restricted this range to be less than $\approx 3\%$ \citep{Margalit2019EoS}. 
This fraction remains an uncertainty which can only be eliminated once the precise EoS is determined. In the following we indicate this quantity with the variable $f_{\rm SNS}$.

Stable NSs produced in NSM are typically very close to their maximum mass.
Hence they will require very little accretion to become supramassive, making them likely to subsequently undergo AICs. This leads to an alternative phenomenology to what discussed earlier on: a short GRB associated with an NSM\footnote{This is under the assumption that NSM leaving behind a stable NS would give rise to a short GRB. However, recent simulations \citep{Ciolfi2020} find that in this situation, while a magnetically-driven collimated jet is launched, it cannot reach relativistic speeds due to high baryonic loading. Hence it may give rise to a different type of transient.}, followed some time later by an AIC at the same location of the short GRB\footnote{Note that, if young magnetars are producing FRBs, then FRB transients would also be produced by post-NSM magnetars  (i.e. prior to the AIC), as suggested by \citet{Margalit2019}}. The range of timescales for this to happen will be discussed in \S~3.3.

\section{Growth of neutron stars in AGN disks}
\label{sec:simulation_ns_agn}

\subsection{Overview of the Model}

To derive the typical properties of AICs and neutron-star mergers in AGN disks, 
we use the model developed in \citet{Tagawa2020,Tagawa2020b}. 
There are several components to this model: 
a supermassive BH (SMBH) at the center of a galaxy, surrounded by an AGN disk and a spherical nuclear star cluster (NSC), which are assumed to be in a fixed steady-state.
For the AGN disk, we employ the model proposed by
\citet[][hereafter TQM]{Thompson2005}. In this model, the disk extends to parsec scales, assuming that it is heated and stabilized by radiation pressure and supernovae from in-situ star formation on parsec scales.
We follow the $N$-body evolution of $\sim 10^5$ individual compact objects (COs, either BHs or NSs) consisting of those in the NSC, those captured inside the AGN disk, and those formed in-situ in the AGN disk. 

In our previous studies (e.g. \citealt{Tagawa2020b}),  
we assumed for simplicity that CO remnants appear instantly at star formation without any delay. Since the finite lifetime of stars is important for NSs, here 
we follow the evolution of in-situ formed stars whose masses are higher than $3\,\Msun$. We assume that stars collapse at the end of their main-sequence lifetime (Appendix~A in \citealt{Tagawa2020}), 
but when the time to collapse becomes longer than the lifetime estimated at the updated mass enhanced due to gas accretion\footnote{Detailed models of star evolution in AGN disks have recently been computed by \citet{Cantiello2020,Dittmann2021,Jermyn2021}.}, we adopt the updated (shorter) lifetime for the  collapse time. 
For a merger condition for stars, we assume the stellar radius to be $R_\odot \left(m_{\mathrm{star}} / \Msun\right)^{1/2}$ \citep[e.g.][]{Torres2010}, where $m_{\mathrm{star}}$ is the stellar mass. When two stars are separated by less than this distance, we assume they merge. 
The initial BH mass is calculated from the mass of a star at collapse through Eq.~(3) in \citet{Tagawa2020} for stars with $\geq20\,\Msun$, while stars with $8$--$20\,\Msun$ form NSs with  mass of $1.3\,\Msun$. 
At collapse, we assign natal kick velocities to supernova-remnant NSs, drawn from a Gaussian distribution with the standard deviation of $265\,\mathrm{km/s}$ for stars with $10$--$20\,\Msun$ (assuming core-collapse SNe) and $20\,\mathrm{km/s}$ for stars with $8$--$10\,\Msun$ (assuming electron capture supernovae; e.g. \citealt{Giacobbo2019}).

CO masses gradually evolve due to gas accretion. We assume that the accretion rate is given by the minimum between a Bondi-Hoyle-Lyttleton rate (modified for a shear flow; see below)  and the Eddington-limited accretion rate with an assumed Eddington ratio ($\Gamma_\mathrm{Edd,NS}$; see \S~3.4 below for more discussion). 
Note that the estimate of the accretion rate is obtained by the capture rate of gas residing within the Bondi-Hoyle-Lyttleton radius, the Hill radius, and also the AGN disk height (Eq.~29 in \citealt{Tagawa2020}). 
NSs are assumed to collapse to BHs at $2.2\,\Msun$\footnote{Note that the precise value of this number depends on the NS EoS, which is not well constrained, as well as on the spin of the NS.}.

Binaries form via several pathways.  
A fraction of COs ($\sim 0.08$ in the fiducial model) is initially in binaries. In the AGN disk, binaries form due to gas dynamical friction during two-body encounters and dynamical interactions during three-body encounters.  We include GW emission, which decreases the binaries' separation and eccentricity during the late stages of their merger.

Interaction with gas in the AGN disk also plays a role in affecting binary evolution: the CO velocities relative to the local AGN disk decrease due to accretion torques and dynamical friction.
Although the binary orbit has been found to expand in some cases  (i.e. for circular, near-equal-mass, prograde binaries embedded in relatively warm circumbinary disks; \citealt{Tang2017, Miranda2017, Munoz2019, Tiede2020, Duffell2020,Heath2020}), a circumbinary disk, fueled by the AGN disk, is likely to decrease the binary's semi-major axis in more realistic thin disks \citep{Tiede2020,Heath2020} and for all retrograde binaries \citep{Nixon2011,Li2021}.
We therefore assume that the semi-major axis and eccentricity further evolve due to type I/II migration torques from a circumbinary disk. COs also migrate toward the SMBH due to type I/II torques from the AGN disk.  We employ standard fitting formulae from the literature for these migration torques \citep{Kanagawa2018}. 

We also account for dynamical interactions with single stars/COs and CO binaries.  The binaries' semi-major axis, velocities, orbital angular momentum directions, and eccentricity evolve due to binary-single interactions as prescribed in \citet{Leigh18}, and the velocities of all COs additionally evolve due to scattering. 

The parameters of the fiducial model are the same as those in \citet{Tagawa2020b}. In this model, the accretion rate onto COs is limited by its own Eddington rate ($\Gamma_\mathrm{Edd,NS}=1$, 
with a radiative efficiency $\eta=0.1$), the gas accretion rate at the outer radius of the simulation ($5\,\mathrm{pc}$) is 0.1 times the Eddington rate (${\dot M}_\mathrm{out}=0.1\,{\dot M}_\mathrm{Edd}$) for the SMBH   with $\eta=0.1$, the SMBH mass is $M_\mathrm{SMBH}=4\times 10^6\,\Msun$, the size of the AGN disk is $r_\mathrm{AGN}=5\,\mathrm{pc}$, and there are initially $5\times 10^4$ NSs within the NSC whose size is $r_\mathrm{CO}=3\,\mathrm{pc}$. 
We assume a radial density profile 
\begin{align}\label{eq:bh_density}
\frac{dN_{\rm NS, ini}(r)}{dr} \propto r ^{\gamma_\mathrm{\rho,NS}}, 
\end{align}
with the slope $\gamma_\mathrm{\rho,NS}=-0.5$ \citep{Hopman2006,Freitag2006}, 
where $N_{\rm NS, ini}(r)$ is the total initial number of NSs within distance $r$ from the central SMBH, 
and a velocity dispersion in the frame comoving with the AGN ($\sigma_{\rm NS} = 0.4 v_\mathrm{kep}(r)$, where $v_\mathrm{kep}(r)$ is the Keplerian velocity at $r$). Note that the above profile we adopted for NSs is shallower, and the velocity dispersion is larger than to those for BHs, as expected from relaxation processes \citep{Hopman2006,Szolgyen2018}. 
We adopt a fiducial disk lifetime of 100 Myr \citep{Martini2004}, but we follow the disk's evolution for up to 1 Gyr.

\begin{figure*}\begin{center}
\includegraphics[width=190mm]{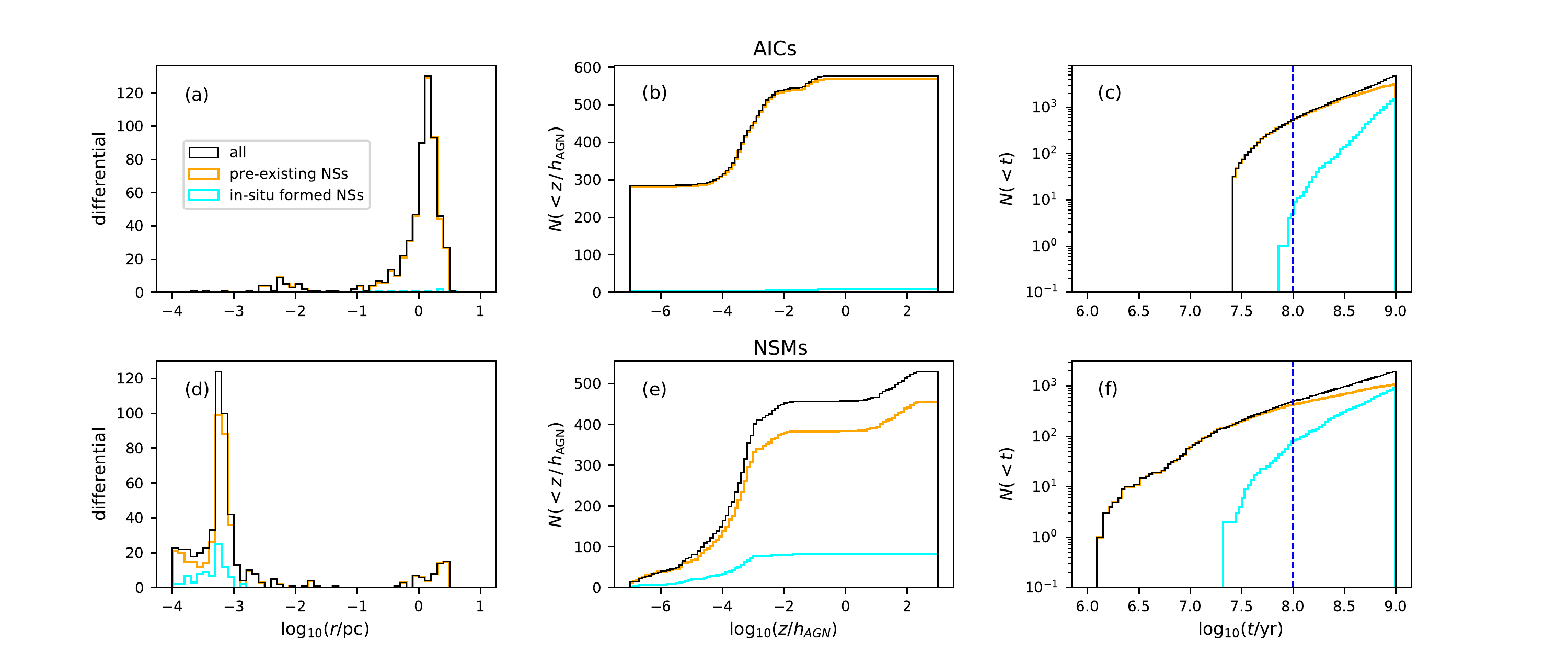}
\caption{
The properties of AICs (upper) and NSMs  (lower) in our fiducial model. 
Left, middle, and right panels show the distribution of the distance of the events from the SMBH at $100\,\mathrm{Myr}$, the cumulative distribution of the orbital height over the scale height of the disk at $100\,\mathrm{Myr}$, and the total number of events up to time $t$, respectively. 
Orange and cyan lines correspond to pre-existing and in-situ formed NSs, respectively and black lines to their sum. The vertical lines in panels~(c) and (f) mark $100\,\mathrm{Myr}$ (the fiducial AGN disk lifetime) for reference. 
}
\label{fig:aic_nsm_prop}
\end{center}\end{figure*}

\begin{figure*}\begin{center}
\includegraphics[width=190mm]{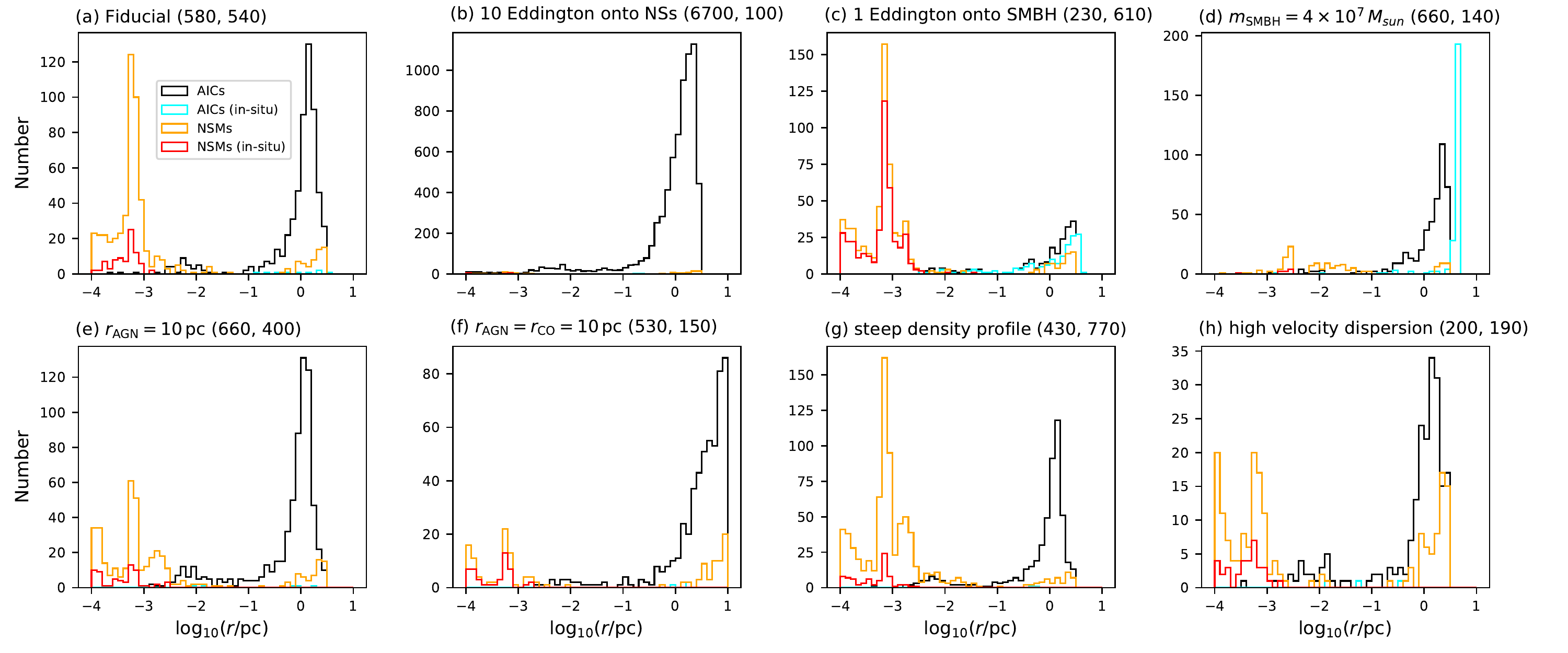}
\caption{The distribution of the distance from the SMBH for AICs and NSMs at $t=100\,\mathrm{Myr}$ in several model variants. 
Black, cyan, orange, and red lines show the results for all AICs, those by in-situ formed NSs, all NSMs, and those by in-situ formed NSs, respectively. 
The left and right numbers in parenthesis are the number of AICs and NSMs, respectively. 
}
\label{fig:aic_nsm_r_dep}
\end{center}\end{figure*}

\subsection{Locations and number of AICs and NSMs per AGN disk}

Fig.~\ref{fig:aic_nsm_prop} shows the properties of AICs and NSMs for our fiducial model.  In total, we find 580 AIC events and 540 NSMs at 100 Myr in this model.
Panel~(a) shows that AICs for pre-existing (orange) and in-situ formed (cyan) NSs 
preferentially occur at around $\sim\,1\,\mathrm{pc}$ from the SMBH. 
Thus AICs are frequent in the outer regions where migration is slower, while in the inner regions of $\lesssim 0.1\,\mathrm{pc}$ NSs quickly migrate to 
gap-forming regions at $r\lesssim 10^{-3}\,\mathrm{pc}$, where they accumulate to a high density, and as a result, NSMs become frequent, driven 
by binary-single interactions (panel {\em d}).

In the fiducial model, the number of AICs from pre-existing NSs and in-situ NSs keeps increasing with time (panel {\em c}). 
In this model BHs are captured in short timescales, while capture of NSs takes a longer time due to both their lower masses and their higher initial velocity dispersion. After being captured into the disk, they are kicked by binary-single interactions, and re-capture takes a longer time for NSs due to strong kicks and weak gas drag on light objects. Additionally, since accretion is limited by the Eddington rate in the fiducial model, mass gain and collapse typically take as long as $\sim40\,\mathrm{Myr}$. A combination of all these effects makes collapse occur more efficiently at later times. This also means that a longer-lived disk will have more AICs. 
The events from in-situ formed NSs are further delayed due to time to collapse and re-capture of NSs kicked by supernovae. 
Since the main-sequence lifetime of an $8$--$20~{\rm M_\odot}$ star is $\sim2$--$20$ Myr, NSMs from stars formed in the disk do not begin until this time (see panel {\rm f}). 
AICs following the formation of stable NSs from NSMs are
expected to occur within a short timescale after the merger, since the NS formed from the merger will likely have a mass close to the threshold for collapse.

All AICs occur in the midplane, deep inside the AGN disk (panel {\em b}), which is a result we verified for several model variations (see below). Similarly, 
most NSMs occur in the AGN disk, while $14\%$ of them are found outside the disk at $100\,\mathrm{Myr}$ (panel {\em e}) due to kicks at binary-single interactions. 
This fraction of mergers outside the disk is reduced to $3\%$ for mergers at $r\leq10^{-3}\,\mathrm{pc}$, where mergers are frequent, 
while $\sim 10\%$ of NSMs at $r\geq \mathrm{pc}$ occur mostly outside the disk. 
The fraction for mergers $r\leq10^{-3}\,\mathrm{pc}$ varies in the range $\sim 3\%$--$11\%$ for models investigated in Fig.~\ref{fig:aic_nsm_r_dep}. 

To give an idea of the robustness of our results, here we examine the dependence of the number of events on several model parameters.  In Fig.~\ref{fig:aic_nsm_r_dep}, we explore the dependence of the results on
a higher maximum accretion rate onto NSs ($\Gamma_\mathrm{Edd,NS}=10$, panel {\em b}), 
a higher gas accretion rate at the outer radius (${\dot M}_\mathrm{out}=1\,{\dot M}_\mathrm{Edd}$, panel {\em c}), 
a higher SMBH mass ($M_\mathrm{SMBH}=4\times 10^7\,\Msun$, panel {\em d}),
a larger AGN disk ($r_\mathrm{AGN}=10\,\mathrm{pc}$, panel {\em e}),
larger sizes for both the NSC ($r_\mathrm{CO}=10\,\mathrm{pc}$) and the AGN disk ($r_\mathrm{AGN}=10\,\mathrm{pc}$, panel {\em f}), 
a steeper radial density profile ($\gamma_\mathrm{\rho,NS}=0$, panel {\em g})
and a higher velocity dispersion ($\sigma_{\rm NS} = 0.8 v_\mathrm{kep}(r)$, panel {\em h}). 
At the top of each panel, we quote, in parenthesis, the total number of AIC events and the total number of NSMs (in round numbers).

As expected, a larger accretion rate onto NSs leads to a larger number of AICs from isolated NSs (6700, instead of the fiducial 580; panel {\em b}). On the other hand, a higher accretion rate leads to a smaller number of NSMs because NSs tend to collapse and become BHs before mergers. Similarly, a larger disk mass associated with a higher accretion rate onto the SMBH (panel {\em c}) or a more massive SMBH (panel {\em d}) is expected to yield more events,
since both gas accretion and migration operate more efficiently in higher gas density environments, facilitating collapse and mergers. 
However, the number of AICs from pre-existing NSs is lower for ${\dot M}_\mathrm{out}={\dot M}_\mathrm{Edd}$. 
This is because the accretion rate onto NSs is limited by the Eddington rate, while the migration speed becomes high in high-density regions, which facilitates NSMs and migration to the inner boundary of the simulation ($r=10^{-4}\,\mathrm{pc}$).
Also, the number of NSMs for a larger SMBH is lower than that in the fiducial model 
since capture of pre-existing NSs is slower for a larger SMBH.  This is because the timescale of capture of NSs to the AGN disk by gas dynamical friction ($t_\mathrm{cap}$) depends strongly on the velocity of NSs relative to the AGN motion ($v_\mathrm{rel}$) as $t_\mathrm{cap}\propto v_\mathrm{rel}^3$.

With the larger size of the AGN disk (panel {\em e}), 
the number of NSMs is somewhat low presumably because the AGN density is lower in this model to stabilize the larger size of the AGN disk, which reduces the inflow rate in inner regions. 
With the larger size of the NSC (panel {\em f}), the number of NSMs is low, because 
pre-existing NSs are initially distributed in outer regions and they take longer time to migrate to the inner regions where mergers are frequent. 
It can be seen that the typical radial distance from the SMBH for pre-existing and in-situ formed NSs 
is related to the size of NSCs and the AGN disk, respectively 
(panels {\em a}, {\em c}, {\em d}, 
and {\em f}). 
For a steeper density profile (panel {\em g}) , 
the number of NSMs is larger, because NSs can efficiently migrate to the inner regions. 
For a higher velocity dispersion (panel {\em h}), the event rate is lower since higher velocities relative to the background AGN disk suppress both captures and accretion. 
In the models with a higher inflow rate, 
due to the high star formation rate to stabilize the disk, the contribution by in-situ formed NSs to the total AIC rate becomes much 
higher (9, 160, and 240 AICs for the models in panels {\em a}, {\em c}, and {\em d}, respectively).

\subsection{Global event rates}

In the following we combine the results from the previous subsection on the event number per AGN, with a cosmological model for the AGN number density, to determine the global rate of events
per unit volume. 

We consider separately the rates for events from pre-existing and in-situ formed NSs as they depend  differently on parameters. 
Following the estimate in \citet{Tagawa2020}, 
the rate density of events from pre-existing NSs is 
\begin{align}\label{eq:event_rate_pre1}
\mathcal{R}_\mathrm{pre} = & \int \frac{\mathrm{d} n_\mathrm{AGN}}{\mathrm{d} M_\mathrm{SMBH}}\frac{f_\mathrm{event} N_\mathrm{NS}}{t_\mathrm{AGN}} \mathrm{d} M_\mathrm{SMBH}\,, 
\end{align}
where $N_\mathrm{NS}$ is the typical number of NSs in NSCs, $t_\mathrm{AGN}$ is the average life time of AGN disks, $n_\mathrm{AGN}$ is the average number density of AGNs in the Universe, and $f_\mathrm{event}$ is the fraction of events per NS in NSCs. 
By using the observed AGN mass function, the observed relations for properties of NSCs and the masses of SMBHs, and assuming that $f_\mathrm{event}$ do not depend on $M_\mathrm{SMBH}$ and events occur for AGNs with the Eddington ratio of more than 0.01, the integral in Eq.~\ref{eq:event_rate_pre1} can be calculated as 
\begin{align}\label{eq:event_rate_pre2}
\mathcal{R}_\mathrm{pre} \sim & 1\,\mathrm{Gpc^{-3}yr^{-1}}\left(\frac{f_\mathrm{event}}{0.02}\right)\left(\frac{t_\mathrm{AGN}}{100\,\mathrm{Myr}}\right)^{-1}\nonumber\\
&\times\left(\frac{\eta_\mathrm{n,NS}}{0.005\,\Msun^{-1}}\right)\,
\end{align}
\citep{Tagawa2020}, where 
the parameter $\eta_\mathrm{n,NS}$ represents the number of NSs per unit stellar mass. 
Due to the mass functions of AGNs and NSCs, the rate is mostly ($\gtrsim 90\%$) contributed by SMBHs in the range $M_\mathrm{SMBH}=10^6$--$10^8\,\Msun$ if $f_\mathrm{event}$ is a constant.  
As the number of NSs for the models in $\S\,\ref{sec:simulation_ns_agn}$ is $5\times 10^4$ and the numbers of AICs and NSMs from pre-existing NSs are 
$\sim 70$--$7000$ and $\sim 100$--$700$, respectively, 
$f_\mathrm{event}$ for AICs and NSMs s are $\sim 0.001$--$0.1$ and $0.002$--$0.01$. 
Note that the number of AICs for a model with 100 and 1000 times the Eddington rate onto NSs is $9.0\times 10^3$ and $1.0\times 10^{4}$, respectively, and thus the number of AICs is limited around $\sim 10^4$, roughly corresponding to the number of NSs captured to AGN disks. 
We assume $\eta_\mathrm{n,NS}$ in the range from $\sim 0.005\,\Msun^{-1}$, corresponding to the Salpeter initial mass function, to $\sim 0.03\,\Msun^{-1}$, reflecting a possible top-heavy initial mass function for NSCs as suggested by \citet{Lu2013}. 
For simplicity, we assume that $f_\mathrm{event}$ is proportional to $t_\mathrm{AGN}$, which is roughly inferred from panels~({\em c}) and ({\em f}) of Fig.~\ref{fig:aic_nsm_prop}. 
With these assumptions and the ranges of values, 
the rates of AICs and NSMs from pre-existing NSs are roughly estimated as 
$\sim 0.07$--$20$ $\mathrm{Gpc^{-3}yr^{-1}}$ and 
$\sim 0.1$--$4$ $\mathrm{Gpc^{-3}yr^{-1}}$, respectively.

Next, we estimate the rate of AICs from in-situ formed NSs. 
Following the estimate in \citet{Stone+2017}, 
the rate density of events from in-situ formed NSs is 
\begin{eqnarray}
    R_\mathrm{IS} &\sim& f_\mathrm{AIC,IS} f_\mathrm{SF/AGN} \eta_\mathrm{n,NS} {\dot \rho}_\mathrm{SMBH}\,,
\end{eqnarray}
where $f_\mathrm{IS}$ is the fraction of events (AICs or NSMs) among in-situ formed NSs, $f_\mathrm{SF/AGN}$ is the star formation rate within the AGN disk over the accretion rate onto the SMBH, and 
${\dot \rho}_\mathrm{SMBH}$ is the total mass accretion rate onto all SMBHs in the local Universe. 
In the models investigated above, 
the fraction of in-situ AICs is 
$2\times 10^{-4} \lsim f_\mathrm{IS} \lsim 6\times 10^{-3}$ 
and that of NS-NS mergers is $10^{-4}\lsim f_\mathrm{IS}\lsim 10^{-2}$. 
Following \citet{Stone+2017}, 
we use $f_\mathrm{SF/AGN}=1$ and 
${\dot \rho}=3\times 10^3\,\Msun \mathrm{Gpc}^{-3}\mathrm{yr}^{-1}$ \citep{Marconi2004}. 
With these values, 
the rate of AICs and NSMs from in-situ formed NSs is estimated as $\sim 0.003-0.5\,\mathrm{Gpc^{-3}yr^{-1}}$ and $\sim 0.002-1\,\mathrm{Gpc^{-3}yr^{-1}}$, respectively. 
Since the contributions from in-situ formed NSs is very small, the total rate of
AICs remains $\sim 0.07$--$20\,\mathrm{Gpc^{-3}yr^{-1}}$, whereas that of NSMs is only slightly increased to $\sim 0.1$--$5\,\mathrm{Gpc^{-3}yr^{-1}}$.
As the rate of NSMs observed by LIGO/Virgo is $\sim 80$--$810\,\mathrm{Gpc^{-3}yr^{-1}}$ \citep{LIGO2020_O3a_Properties}, this implies that 
between $\sim 0.01$\% and $\sim 6$\%
of NSMss can potentially originate from AGN disks, while AICs can be an order of magnitude more common.

Note that we calculate the rates using the observed AGN mass function and the growth rate of SMBHs for low redshift \citep{Greene2007,Marconi2004}. 
On the other hand, as the AGN density increases with the redshift, the event rates are likely to be enhanced by a factor of $\sim 2$ at redshift $z=1$ \citep{Yang2020_redshift}.

\subsection{Potential EM emission from AICs in AGN disks}

As discussed in \S2, strong electromagnetic signatures from AICs require either a substantial amount of mass, or a strong  magnetic field, to surround the NS at the time of its collapse. 
In this section, we will discuss these two possibilities in turn.

First, to estimate how much mass might be
surrounding the NS at its collapse, it is useful to start by considering the gas mass bound to the NS at a given time.  In our fiducial model with a SMBH of
$4\times 10^6 M_\odot$, AICs are found to be concentrated at a distance of about 1~pc from the central SMBH.
Gas is bound to the NS 
approximately within the "accretion radius",
\begin{equation}
R_{\rm acc}=\frac{GM_{\rm NS}}{\sigma^2}\,, 
\end{equation}
with $\sigma^2\equiv (r_{\rm H}\Omega)^2 + c_s^2$
\citep[see, e.g.][]{Stone+2017,Tagawa2020}.  Here $M_{\rm NS}$ is the mass of the NS, $c_s$ is the sound speed of the nearby disk gas, and $r_{\rm H}=r(M_{\rm NS}/3M_{\rm SMBH})^{1/3}$ is the Hill radius of a NS on a circular orbit around the SMBH at distance $r$ with Keplerian orbital frequency $\Omega$.  This accretion radius is 
roughly the smaller between the traditional thermal Bondi radius $r_{\rm B}=GM_{\rm NS}/c_s^2$ and the 
Hill radius, requiring that neither thermal motions, nor shear motions, should unbind the gas from the NS.

In the TQM model, for our fiducial parameter choices with a $4\times10^6~{\rm M_\odot}$ SMBH, at $1\,$pc from the nucleus, 
we find $c_s/v_{\rm K}=H/r\approx 0.003$ (where 
$v_{\rm K}\approx 130~{\rm km~s^{-1}}$ 
is the Keplerian gas velocity of the background AGN disk and $H$ is the disk scale height), 
$r_{\rm H}\approx 0.005$pc and $r_{\rm B}\approx 0.03$pc.  
The mass density at this location is 
$\rho\sim 10^{-16}$~g~cm$^{-3}$ 
and the surface density is $2H\rho\sim 
2~{\rm g~cm^{-2}}$,
implying that the mass inside the accretion radius at any given time is negligibly small ($\sim 10^{15}g$). Furthermore, the inflow time $\sim r_{\rm H}/c_s$ from this distance to the NS is as long as $\sim 10^4$~yr.   

On the other hand, we need to consider how much mass may have {\em accumulated} over time closer to the NS by the time of its collapse.
The accretion rate $\dot{M}=\rho\sigma (2 R_{\rm H})(2H) = {\rm few}\,\times10^{-5}~{\rm M_\odot yr^{-1}}$ implies that
if this rate persisted on small scales, all the way down near the NS, it would accumulate $\sim10^3~{\rm M_\odot}$ over the disk lifetime of a few $\times$ 10 Myr.   However, only a small fraction of this nominal large-scale accretion rate is likely to reach the vicinity of the NS, with the rest either accumulating in a massive envelope, resembling a Thorne-$\dot{\rm Z}$ytkow object (\citealt{Thorne1977}; see also recent work in the context of AGN disks by \citealt{Wang2021}), and/or ejected in outflows.

Since $r_{\rm H}\gsim H$, the NS marginally satisfies the thermal gap-opening condition.  On the other hand, with $\alpha=0.1$, it marginally fails the viscous condition, and the expected reduction in the surface density within the annulus extending over the Hill radius is only a factor of a few (see eq. 10 in \citealt{Kanagawa2018}).  Since $r_{\rm B}>r_{\rm H}$, accretion towards the NS would occur from gas streams librating on horse-shoe orbits around the orbital radius of the NS, facilitated by shocks near the Hill radius.  The situation is analogous to planetary accretion, which typically finds a circumplanetary disk inside the planet's Hill radius (e.g. \citealt{Szulagyi2014}).  This suggests that accretion towards the NS would be disk-like on large scales (although this is not fully understood; e.g. \citealt{Derdzinski+2019} find that the captured gas has little angular momentum and forms a quasi-spherical cloud).

The rate $\dot{M} = {\rm few}\,\times10^{-5}~{\rm M_\odot yr^{-1}}$  is roughly three orders of magnitude above the NS's Eddington rate, and radiation therefore likely impacts the accretion geometry and the fraction of gas reaching the NS.
While super-Eddington accretion rates onto compact objects have been shown to be possible \citep{Jiang+2014}, this assumes that this fueling rate is delivered close (within tens of gravitational radii) of the central object.   In this region, approximately half of the mass reaches the compact object, with the rest emerging in vertical outflows or a radiatively-driven polar outflow in the innermost regions (\citealt{Jiang+2014}, see also \citealt{Narayan1995, Igumenchev2000,Proga2003}). 
 If not all the mass  is ejected to infinity, but a fraction remains bound and/or collides with infalling mass further out and falls back, it could end up accumulating a compact mini-disk near the NS. While the size and amount of mass in this minidisk is uncertain, it is some small fraction of the total $\sim 10^3~{\rm M_\odot}$ being fed to the NS from larger scales over several 10 Myr. 
For a phase of rapid accretion post-collapse to set in, a certain amount of mass
is needed within a few tens of gravitational radii from  the collapsing NS. 
In the GRMHD simulations by \citet{Giacomazzo2012} the NS was surrounded by  a torus of $\sim 0.1M_\odot$ and accretion onto the BH, triggered by spacetime dynamics, led to accretion rates of $\sim 10^{-2}M_\odot$~s$^{-1}$, similar to those inferred for short GRBs. However, more generally, if a mini disk of mass $M_d$ surrounds the newly formed BH at a radius $R_d$, a phase of accretion at the rate $\dot{M}_{d}\sim M_d/t_{\rm visc}(R_d)$ will ensue due to viscous dissipation, where $t_{\rm visc}$ is the viscous timescale \citep{Shakura1973}.
  Therefore, longer and less bright transients are possible if less mass near the collapsing NS is accumulated and/or the accumulated mass is spread over a more extended region.

Unlike GRB-like events, FRB-like transients do not directly depend on the environment of the AGN. However, the strength
of their emission indirectly does since the NS magnetic field decays with time. The amount of decay depends on several factors, such as the relative fraction of dipolar versus toroidal field, and the fraction of the field in the (superconducting) core. Magnetothermal simulations by \citet{Vigano2013} found that a dipolar field of $10^{14}$~G can decay by a factor varying between 5\% to one order of magnitude depending on the above conditions in the NS interior. 
Our results have shown that within a $10^8$~yr AGN lifetime 
about 580 NSs undergo AICs in the fiducial model. 
The typical time per NS to accrete a critical amount of mass for collapse is 30~Myr in our 
simple fiducial model in which all the NSs start with a mass of $1.3M_\odot$ and undergo AIC at $2.2~M_\odot$. In reality, there will be a distribution of timescales reflecting the distribution of NS masses at birth. Using 
30~Myr as a representative timescale, we hence note that a magnetar field may substantially decay over this time, leading to a more typical NS field at the time of the AIC. 
However, the FRB-type AICs predicted by the model of \citet{Falcke2014}
can have luminosities comparable to those of the observed FRBs for typical NS fields.
For the observed population, by fitting with a Schechter function,
\citet{Luo2020} determined a lower cut-off in luminosity $\sim 10^{42}$~erg~s$^{-1}$. This can be achieved with magnetic fields down to $B\sim$~a few~$\time 10^{11}$~G. Hence, even accounting for considerable field decay during the accreting phase of the NS, 
 the rate of FRB-like events from isolated NSs
will be comparable to the sum of the rates of AICs from pre-existing and from in-situ formed NS, but with a range of intensities which depends on both the magnetic field at birth as well as on the time lapse between birth and the AIC.

For AICs produced following NSM events, the timescale for accretion is expected to be much shorter, since the post-merger NS is already very massive, and hence magnetic field decay is not expected to be important. The rate of these events is  given by $f_{\rm SNS} N_{\rm NSM}$; they consist of a short GRB at the time of the NSM, and an AIC sometime later.  The time lapse between the short GRB and the FRB depends on how close the mass of the post-merger NS is to the maximum allowed mass. Hence we expect a distribution of timescales. 
As a reference example, consider a 
NS whose mass is $\sim 10^{-6}M_\odot$ below the maximum mass for collapse.  If this
NS accretes at 10 times the Eddington rate, it will take $\sim~1$~year to grow by this deficit, and the delay between the short GRB and the FRB will be $\sim~1$~year. 

\section{Observability of Electromagnetic  transients from AIC's in AGN disks}

As discussed in \S~2 and \S~3.4, AICs can give rise to different types of electromagnetic transients, depending on the circumstances surrounding the collapse. If at the time of collapse the NS is surrounded by a dense 
torus, then, upon collapse to a BH, a brief phase of very rapid accretion would likely give rise to a 'short GRB-like' type of transient. These are characterized by a very brief (generally sub-second) $\gamma$-ray emission, followed on much longer timescales by longer-wavelength emission spanning the entire electromagnetic spectrum, from the X-rays to the radio band.
Distinguishing such transients from the standard short GRBs resulting from an NS-NS merger and possibly a BH-NS merger, would need to rely on the simultaneous detection of a gravitational wave counterpart. The lack of detection of a NS-NS or NS-BH GW signal within the LIGO/Virgo horizon for a short GRB type of event associated with an AGN disk would make the event a strong candidate for an AIC. 

Given that AIC events can be distinguished from  binary mergers with the help of potential GW counterparts, the
next question is how the dense environment of an AGN disk affects the observability of GRB-like transients. 
This question was recently addressed by \citet{Perna2021}. The observational appearance of the transient is found to depend on its location (and hence the local conditions) within the disk. Under the simplifying assumption of full ionization in the disk, they found that the main determinant of the appearance of a GRB transient is the time that the emitting jet takes to pierce
the disk photosphere (to electron scattering): radiation emitted afterwards encounters an optically thin medium, and hence it escapes unaltered, whereas radiation emitted from inside the disk photosphere 
emerges on the diffusion timescale. 
In the following, we will specialize the observability discussion 
based on the typical location of the AICs that we found in Sec.~3 for our fiducial model, and on the specifics of the opacity at that location based on the TQM disk model.  

We have shown (\S~3) that AICs by pre-existing NSs are concentrated at a distance of about 1~pc, 
corresponding to $\sim$~half the size of the NSCs \citep{Gultekin2009}. 
The disk is neutral in those regions  if unaffected by sources of radiation; therefore its opacity at the time of the AIC is sensitively dependent on the pre-AIC ionization history. Generally speaking, a medium is kept ionized if the ionization time is comparable to the recombination time. At the distance of 1~pc in our fiducial TQM disk model, the timescale for recombination is  
\begin{equation}
t_{\rm rec}\sim 0.7 ~\left(\frac{T}{10^4{\rm K}}\right)^{1/2}\left(\frac{n_e}{6\times 10^{7}{\rm cm}^{-3}}\right)^{-1}~{\rm days}, 
\end{equation}
having used the local density $n_e$ at 1~pc and the temperature $T\approx 10^4$K for photoionized gas (using the case-B recombination coefficient $\alpha_B=2.6\times10^{-13} {\rm cm^3~s^{-1}}$). As an order of magnitude estimate for the ionization time, let us consider for simplicity the case of pure Hydrogen, and a photoionizing flux $L_0/(4\pi H^2)=\nu_0L_{\nu_0}/(4\pi H^2)$ at the ionization frequency $\nu_0$. The photoionization time is then given by 
\begin{equation}
 t_{\rm ion} \sim \frac{ h_{\rm P}4\pi  H^2}{L_{\nu_0} \sigma_0}\,, 
\end{equation}
where $\sigma_0=6.3\times10^{-18}~{\rm cm^2}$ is the cross-section at the ionization threshold $h_{\rm P}\nu_0=13.6$eV (here $h_{\rm P}$ is the Planck constant). In order for gas to remain ionized, that is $t_{\rm ion}\lesssim t_{\rm rec}$, the requirement is for the ionizing luminosity to be $L_{\rm \nu_0}\gtrsim 1.9\times 10^{17}$~erg~s$^{-1}$~Hz$^{-1}$. 
In order to evaluate how this compares with the potential output from an NS accreting at about the Eddington rate, we take as an illustrative spectrum that of ultra-luminous X-ray sources (ULXs), since several ULXs have been associated with accreting NSs \citep{Bachetti2014}. The analysis of several sources by \citet{Koliopanos2017} showed that a generic feature is a double multicolor blackbody, with the hotter component at a temperature $kT\gtrsim 1$~keV and the cooler one, associated to a disk truncated at the NS magnetosphere, at $kT\lesssim 0.7$~keV. As an illustrative example, let us consider a blackbody with peak luminosity $\nu L_\nu \sim 10^{38}$~ers~s$^{-1}$ at 1~keV. This implies  $L_{\nu}\sim 4\times 10^{20}$~erg~s$^{-1}$~Hz$^{-1}$ at 1~keV. The ionizing luminosity at 13.6~eV is then on the order of  $8\times 10^{16}$~erg~s$^{-1}$~Hz$^{-1}$ (simply using the Raleigh Jeans scaling down to lower energies),  comparable to the value required for $t_{\rm ion} \sim t_{\rm rec}$, hence
making the results sensitively dependent on the ionizing spectrum of the accreting NS.
Additionally, radiation from magnetized NSs accreting at high rates is expected to have some degree of beaming: the accretion column focuses the emission in a 'pencil' fashion \citep{Basko1975}. Once the medium becomes optically thick at
accretion rates above $\sim 10^{37}$~erg~s$^{-1}$, radiation starts to escape along the sides of the column, creating a 'fan' beaming \citep{Schonherr2007}. In our context, this means that there may be lines of sights within the AGN disk more affected by the radiation than others. 

Given the above uncertainties, in the following we will discuss observability 
based on two extremes: (a) medium unaffected  
by the radiation of the accreting NS (at least along the line of sight to the AIC transient); (b) medium fully ionized by the accretion-generated radiation.

Let us first consider case (a), in which the medium along the line of sight to the transient is initially neutral.
The prompt $\gamma$-rays are expected to escape unhindered.
With a surface density $\Sigma \sim 1$~g~cm$^{-2}$ from midplane at the $1$~pc location
(in our fiducial model), the radiation remains optically thin down to $E\gtrsim 6-7$~keV (see e.g. Fig.~1 in \citealt{Wilms2000}, corresponding to an absorbing column with the metallicity in the interstellar medium),
whereas the longer-wavelength radiation encounters an optically thick medium.
However, 'GRB-like' transients are typically characterized by early X-ray/UV radiation which is very effective at photoionizing  gas along the line of sight and at sublimating dust
\citep{Waxman2000,Perna2003dust}, hence electron scattering is going to be the main source of opacity for the longer-wavelength radiation arriving at later times. Since the medium is optically thin to Thomson scattering ($\tau\sim 0.4$ at $\sim 1$pc), this radiation is not expected to be subject to significant diffusion. Given that the flux level needed to keep the gas ionized is quite low (as determined earlier in this section), we expect that, even when the gas starts neutral, a GRB-like transient will emerge largely unaffected. Additionally,  
since dust is expected to be destroyed promptly, the optical and infrared radiation should not suffer appreciable dimming. This optical/IR radiation, when the prompt $\gamma$-rays are missed due to beaming effects and/or no $\gamma$-ray detector pointing in their direction, would then appear as 'orphan afterglows' and should be searched for among the unidentified AGN variability.   The luminosity of a typical, non-diffused GRB afterglow would generally outshine the typical luminosity of an AGN, especially since afterglows in dense media evolve faster and are initially brighter than those in the interstellar medium (see discussion in \citealt{Perna2021}).  

As discussed in Sec.~2, while a GRB-like transient can be produced by post-AIC accretion onto the BH, an FRB-like transient is expected from the disrupted magnetosphere during the collapse, and hence {\em prior} to any GRB emission. As such, if the AGN medium is neutral at its location, the pulse is not expected to be affected by either dispersion nor free-free absorption. Hence the predicted phenomenology would be that of an FRB-like transient promptly followed (on a $\sim $~sec timescale) by a short GRB-like transient.
We note however, that it would be unlikely for an observer to detect both the FRB and the GRB. Infact, the EM radiation from the the collapsing NS is radiated most efficiently at angles $\pm 50^\circ$ from the dipolar axis \citep{Lehner2012}, whereas a GRB jet is expected to emerge in the plane perpendicular to that of the accretion disk, and hence uncorrelated with the above. From an observational point of view, these events should hence be searched among FRBs followed, on a timescale of days to weeks, by
orphan afterglows in the optical/IR.

Let us consider next the opposite extreme of case b), in which the medium (at least along the line of sight to the observer) has been fully ionized by the accreting NS prior to collapse. In this case a GRB-like transient  and its X-ray/optical afterglow would emerge largely unaffected since the disk is optically thin to Thomson scattering, as discussed above. However, the situation would be very different for an FRB-like transient. The dispersion measure from the disk mid-plane at 1pc is
\begin{eqnarray}
{\rm DM} = \int_0^\infty n_e(z) dz=
\sqrt{\frac{\pi}{2}}\frac{\rho_0}{m_p} H \sim \nonumber \\
 2\times 10^{5} \left(\frac{\rho_0}{10^{-16}{\rm g}~{\rm cm}^{-3}} \right)\left(\frac{H}{0.003{\rm pc}}\right)\,{\rm pc}~{\rm cm}^{-3}\,
    \label{eq:DM}
\end{eqnarray}
while the optical depth to free-free absorption is
(using the approximation by \citealt{Mezger1967}),
\begin{eqnarray}
\tau_{\rm ff} &\approx& 3\times 10^{6} \left(\frac{T}{ 10^4{\rm K}}\right)^{-1.35}\left(\frac{\nu}{\rm GHz}\right)^{-2.1}\nonumber \\
&\times& \left(\frac{\rm EM}{9\times 10^{12}~{\rm pc}~{\rm cm}^{-6}}\right)\,,
    \label{eq:tauff}
\end{eqnarray}
where $n_e(z)$ is the electron density along the line of sight to the observer within the AGN disk, and $EM= \int_0^\infty n^2_e(z) dz$
 is the emission measure evaluated 
with the local disk conditions; the assumed temperature is as typical of photoionized gas. 
It is evident that, under the most conservative conditions that the accreting NS fully ionizes the 
medium along the line of sight to the observer and does not affect the medium otherwise, 
FRB-type of transients will not be observable.
However, we note that radiation-driven outflows during the high-rate accretion phase of the NS \citep{Oshiga2005,Jiang+2014} can potentially carve a bubble in the disk, as shown by \citet{Kimura2021}, hence reducing the density of the medium and facilitating the emergence of the radio emission from the transients.
 
For FRB-type events produced by post-NSM magnetars, the situation is more complex and dependent on the radial location within the disk. 
About $10\%$ of NSMs occur at large disk radii $\sim~1$~pc. All of these events are found above  several disk scale heights (see panel (e) of Fig.1).
Even under the most pessimistic circumstances of the gas along the line of sight to the observer having been fully  ionized by the accreting NS prior to collapse, the optical depth to free-free absorption becomes smaller than 1 for $z\gtrsim 3.6 H$, and the DM from that location is $\sim 70$~pc~cm$^{-3}$. Hence we expect these post-NSM FRBs to be generally detectable. 

On the other hand,  for the majority ($\sim 90\%$) of NSM events which occur at much smaller radii
(around $\sim 10^{-3}$~pc), we need a more
detailed analysis.
The gas at those locations is ionized, and hence absorption can pose a significant observational challenge. The most pessimistic case is the one in which the FRB sites are in the mid-plane, and there has been no prior disruption of the disk gas due to the binary NS merger. The dispersion measure from the midplane at $\sim 10^{-3}$~pc is prohibitively high, DM$\sim 6\times 10^{9}$~pc~cm$^{-3}$, and so is the optical depth to free-free absorption, $\tau_{\rm ff}\sim 10^{18}$. 
As a reference, note that CHIME/FRB detects transients up to a dispersion measure of $\sim 13,000$~pc~cm$^{-3}$ \citep{CHIME2018}. With $\rho \propto e^{-z^2/2H^2}$, the DM enters in the observable range for $z\gtrsim 4.2 H$, while the optical depth to free-free absorption becomes smaller than 1 for $z\gtrsim 3.9 H$. 
In our fiducial model, only about 3\% of NSMs occur above a few scale heights, and hence could in principle be observable with CHIME/FRB and be optically thin to absorption. For the bulk of FRB-like events following NSMs occurring at $\lesssim 10^{-3}$~pc, the observational prospective is largely dependent on the effect of the prior NSM on the disk environment. \citet{Zhu2021} examined the evolution of a jet emerging from NSMs located in the migration traps of an AGN disk, where the bulk of our events also occurs \citep[but see][]{Kocsis2011,Pan2021}. 
They find that the jet energy is deposited within the disk material to power a hot cocoon, and that this cocoon is energetic enough to break out from the AGN disk. While it is likely that the jet propagation leaves behind a lower density funnel, the detailed modification of the gas density profile in the disk (and hence the corresponding dispersion and attenuation of an FRB-like signal following the NSM)  can only be computed via dedicated numerical simulations, and will be reserved to a future investigation (Lazzati et al. 2021, in prep).
 However, we note that the recent work by \citet{Kimura2021} has already argued how, in AGN disks,  accretion onto a circumbinary disk surrounding the compact object binary will generate a strong radiation-driven outflow which then carves a low-density bubble. Such an environment prior to the AIC would largely enhance its detection probability.

\section{Summary} 

The accretion-induced collapse (AIC) of a neutron star to a black hole is a well studied phenomenon from a numerical point of view, and it has been suggested as the engine for astrophysical phenomena such as 
both long and 
short GRBs and FRBs. 

Here we have discussed how the disks of AGNs provide a natural environment for this phenomenon to occur, and we have quantified its occurrence rate, considering both accreting isolated NSs (whether pre-existing or formed in-situ) and stable NSs produced in NS-NS mergers.
Over a range of models we explored, we found that the rate from the former is $\sim 0.07-20$~Gpc$^{-3}$~yr$^{-1}$, while the latter is $\sim f_{\rm stable}$(0.5-5)~Gpc$^{-3}$~yr$^{-1}$,
where $f_{\rm stable}$ is the fraction of NS-NS mergers which leave behind a stable NS, which depends sensitively on the equation of state of the NS.

Electromagnetic radiation associated with AICs can be powered by accretion from a mini-disk surrounding the NS prior to collapse, and/or by reconnection in the magnetic field outside the horizon post collapse of the NS.  The former events would have features more in common with those of short GRBs, while the latter more similar to those of the FRBs. 

 Detectability is strongly dependent on the location of the transients, as well as on the ionization and dynamical pre-AIC history of the medium. AICs from isolated NSs occur at around 1~pc distance from the SMBH and are concentrated in the mid-plane. The AGN disk is cold and neutral at those distances, which would allow FRBs to not suffer significant dispersion nor attenuation.
However, if the line of sight to the observer has been significantly ionized by the radiation from the accreting NS prior to collapse, then a large dispersion measure and a large optical depth to free-free absorption would hinder detection, unless outflows by the accreting NSs have also excavated a low-density bubble. On the other hand GRB-like transients from those outer locations would escape mostly unaffected, since the early radiation is very effective at ionizing, and the disk is optically thin to Thomson scattering in those regions.
When both transients are detectable and beamed in the same direction to the observer, these AICs would appear as FRBs followed ($\sim$~a few seconds timescale) by short GRBs.  However, we noted that
the $\gamma$-ray emitting jet will most likely point away from the radio pulse, hence
making the detection of both an FRB and a short GRB not very likely. Neverthless, since 
the longer wavelength afterglow (optical/IR) is largely isotropic, these AIC events, if an FRB is detected,  would be followed by optical and IR radiation; this should be searched for among AGN variability.

A subset of AICs is expected to be produced
following NSMs which leave behind stable NSs.
FRB-like transients associated with the fraction ($\sim 10\%$) of these which occur at several scale heights above the disk 
are expected to be detectable, whereas the majority which occur inner in the disk at $\sim 10^{-3}$~pc will be heavily obscured unless the preceding NSM excavates a low density bubble. 

To conclude, we have highlighted a new, potentially rich phenomenology in AGN disks, whose observability is tightly coupled with the disk conditions and with the interplay between compact object evolution and the AGN disk itself. As the future years are expected to bring a large wealth of data on FRBs with CHIME/FRB \citep{CHIME2018} and optical/IR observations with the Vera Rubin Observatory\footnote{https://www.lsst.org}, we hope that this work, together with observational constraints on the predicted rates, will help us gain a further inside into the conditions of the AGN disks and the compact objects that live in them.  
\\

\bigskip
\bigskip

\acknowledgments
We thank Brian Metzger for very useful comments on our manuscript.
The Flatiron Institute is supported by the Simons Foundation. RP acknowledges support by NSF award AST-2006839 and from NASA (Fermi) award 80NSSC20K1570.
HT acknowledges support by the Grants-in-Aid for Basic Research by the Ministry of Education, Science and Culture of Japan (HT:17H01102, 17H06360). 
ZH was supported by NASA grant NNX15AB19G and NSF grants AST-2006176 and AST-1715661.
Simulations were carried out on Cray XC50 
at the Center for Computational Astrophysics, National Astronomical Observatory of Japan.  IB acknowledges the support of the Alfred P. Sloan Foundation.

\bibliographystyle{aasjournal}
\bibliography{refs}

\end{document}